\documentclass[]{ceurart}

\usepackage{listings}
\usepackage{url}
\hypersetup{
    colorlinks=true,
    linkcolor=blue,
    filecolor=magenta,      
    urlcolor=blue,
    pdfpagemode=FullScreen,
    }
\usepackage{tabularx}
\newcolumntype{t}{>{\hsize=23\hsize}X}
\newcolumntype{a}{>{\hsize=4\hsize}X}

\usepackage{subcaption}
\usepackage{todonotes}

\begin{document}

\copyrightyear{2024}
\copyrightclause{Copyright for this report is held by its authors.
  Use is permitted under Creative Commons License Attribution 4.0
  International (CC BY 4.0).}

\conference{ }

\author[1]{Luca Rossetto}[%
    orcid=0000-0002-5389-9465,
    email=luca.rossetto@dcu.ie]

\author[2]{Klaus Schoeffmann}[%
    orcid=0000-0002-9218-1704,
    email=ks@itec.aau.at]

\author[1]{Cathal Gurrin}[%
    orcid=0000-0002-9218-1704,
    email=cathal.gurrin@dcu.ie]  

\author[3]{Jakub Lokoč}[%
    orcid=0000-0002-3558-4144,
    email=jakub.lokoc@matfyz.cuni.cz] 

\author[4]{Werner Bailer}[%
    orcid=0000-0003-2442-4900,
    email=werner.bailer@joanneum.at]

\address[1]{Adapt Centre, School of Computing, Dublin City University, Dublin, Ireland}
\address[2]{Institute of Information Technology, Klagenfurt University, Austria}
\address[3]{Department of Software Engineering, Charles University, Prague, Czechia}
\address[4]{DIGITAL -- Institute for Digital Technologies, JOANNEUM RESEARCH, Graz, Austria}

\begin{abstract}
    This report presents the results of the 13\textsuperscript{th} Video Browser Showdown, held at the 2024 International Conference on Multimedia Modeling on the 29\textsuperscript{th} of January 2024 in Amsterdam, the Netherlands.
\end{abstract}

\title{Results of the 2024 Video Browser Showdown}

\maketitle

\section{Introduction}

The Video Browser Showdown (VBS) is an annual interactive video retrieval competition held at the International Conference on MultiMedia Modeling (MMM).
This report presents the results of the 13\textsuperscript{th} edition of this international competition (VBS 2024), which took place on  29\textsuperscript{th} of January 2024 in Amsterdam in the Netherlands.
It provides an overview of the participating teams and visualizes the scores with respect to teams and task types.
The aim of this report is to summarize the results without providing an in-depth discussion.
For a more detailed discussion on past instances of the VBS, see the detailed analysis papers from 2023~\cite{DBLP:journals/access/VadicamoABCGHLLLMMNPRSSSTV24}, 2022~\cite{DBLP:journals/mms/LokocABDGMMNPRSSSKSVV23}, 2021~\cite{DBLP:journals/ijmir/HellerGBG0LLMPR22}, etc.
The data presented in this report was generated by the Distributed Retrieval Evaluation Server~\cite{DRES-TOMM}, which is used as a central coordination and evaluation infrastructure for VBS.
The raw data exports are available from the VBS archive.\footnote{\url{https://github.com/lucaro/VBS-Archive}}

\section{Competition setup}

As in the previous year~\cite{DBLP:journals/access/VadicamoABCGHLLLMMNPRSSSTV24}, VBS 2024 used challenging datasets -- two shards of the V3C collection \cite{RossettoSAB19} with 2,300 hours of heterogeneous video content from 17,235 video files downloaded from Vimeo, and the small MVK dataset \cite{MVK} with challenging 12 hours of homogeneous video content from 1,374 marine videos. For the first time at VBS, these two datasets were complemented by another highly challenging dataset from the medical domain with about 100 hours of video content from 72 gynecological laparoscopies.

Expert users solved four different task types: known-item search (KIS) tasks that can be presented with a textual hint (KIST) or a visual hint (KISV), ad-hoc video search (AVS), and question-answering (QAS) tasks. Visual KIS tasks were presented as a scene played in a loop, while during textual KIS and AVS tasks, a text describing the searched scene/concepts was presented. QAS tasks were presented as a combination of textual description and video clips (e.g., ``\textit{How many nights are shown in this video?}''). Unlike expert users, novice users tried only three task types, skipping textual KIS tasks. The list of task types, their duration and presentation form is presented in Table \ref{tab:taskTypes}. 
These different tasks cover different combinations of properties such as search need presentation, number of relevant requested items, and search need presentation quality (i.e., level of detail) in the space of possible task categories~\cite{DBLP:conf/mmm/LokocBBGHJPRSVV22}.

QAS tasks were used for the first time in VBS in the 2024 edition. They share some properties with KIS tasks, as the search need is also presented as text and video clip, and there is a single correct answer. The video hint serves to set the context of a video in which the question is to be answered. While it would be desirable to ask a text-only question against the entire dataset, this is a simplification that avoids guaranteeing the uniqueness of the answer in this large collection. Additionally, QAS tasks require efficient browsing features, as the answer may require inspecting a time-point in the video far from the clip shown in the query or even the combination of hints appearing throughout the video. 

\begin{figure}[!ht]
    \centering
    \includegraphics[width=0.925\linewidth]{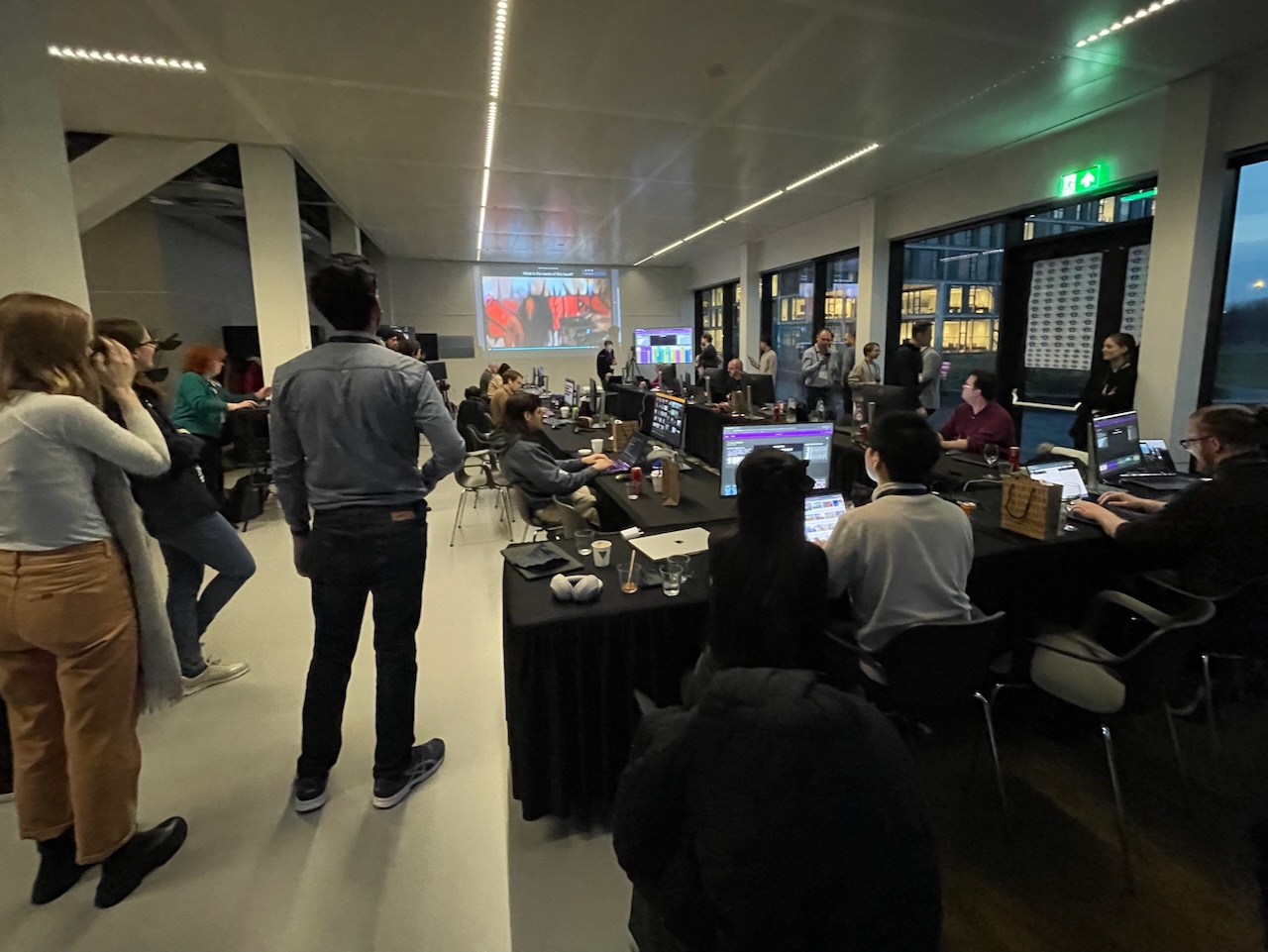}
    \caption{Setup of VBS 2024: a u-shape arrangement of teams around the projector showing the evaluation server display (tasks and results).}
    \label{fig:vbs2024setup}
\end{figure}

\begin{figure}[!ht]
    \centering
    \includegraphics[width=0.925\linewidth]{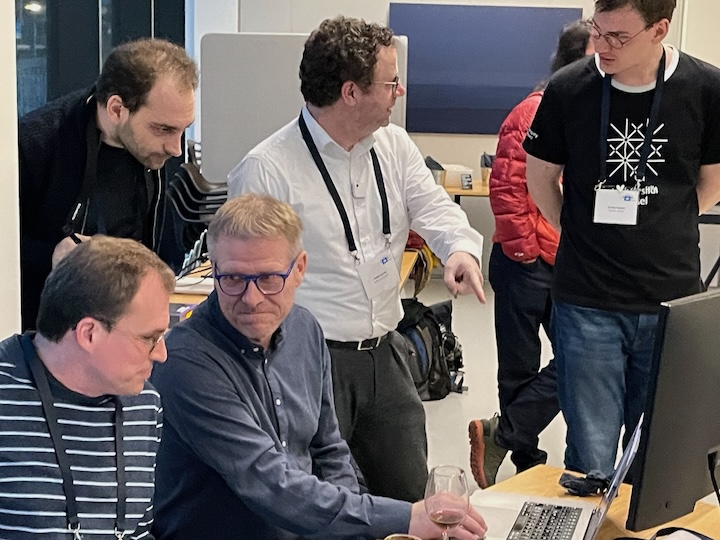}
    \caption{Live judging of submissions by a team of on-site and offline judges for tasks without (complete) ground-truth.}
    \label{fig:vbs2024judges}
\end{figure}

The tasks were presented through a projector and large displays directly connected to the evaluation server (DRES)~\cite{DRES-TOMM}. Teams were in a U-shaped arrangement so that they could clearly see the evaluation server screen, but less so the system and solutions of other teams (see Figure~\ref{fig:vbs2024setup}). Due to the huge dataset, the ground-truth of AVS tasks with multiple answers cannot be pre-annotated and automatically checked. In addition, answers to QAS tasks are often in different languages and syntaxes. Therefore, VBS 2024 used live judges who manually assessed submissions that could not be automatically checked during the competition (see Figure~\ref{fig:vbs2024judges}). 11 researchers served as live judges during at least one of the sessions.

\begin{table}[]
    \centering
    \begin{tabular}{l|l|l}
        type & duration & task presentation \\
        \hline
        visual known-item search (KISV) & 5 minutes & scene in the loop  \\
        textual known-item search (KIST) & 7 minutes & text revealed incrementally  \\
        ad-hoc video search (AVS) & 5 minutes & short text description \\
        questions answering (QAS) & 5 minutes & scene in the loop and text question 
    \end{tabular}
    \caption{Task types at VBS 2024}
    \label{tab:taskTypes}
\end{table}

\pagebreak
\section{Teams}

In 2024, 12 teams participated in VBS. In contrast to previous years, team members were scored individually rather than in aggregate.
The following lists the participating teams, ordered by their final rank in the evaluation.

\begin{enumerate}
    \item VISIONE~\cite{DBLP:conf/mmm/AmatoBCFGMVV24}
    \item Vibro~\cite{DBLP:conf/mmm/SchallHBJ24}
    \item PraK~\cite{DBLP:conf/mmm/LokocVSBS24}
    \item ViewsInsight~\cite{DBLP:conf/mmm/VuongHNTLPNGT24}
    \item diveXplore~\cite{DBLP:conf/mmm/SchoeffmannN24}
    \item TalkSee~\cite{DBLP:conf/mmm/GuWHSWL24}
    \item VIREO~\cite{DBLP:conf/mmm/MaWN24}
    \item vitrivr-VR~\cite{DBLP:conf/mmm/SpiessRS24}
    \item vitrivr~\cite{DBLP:conf/mmm/GasserAFSWR24}
    \item VERGE~\cite{DBLP:conf/mmm/PantelidisPGASM24}
    \item VideoCLIP~\cite{DBLP:conf/mmm/NguyenQHNG24}
    \item Exquisitor~\cite{DBLP:conf/mmm/KhanZSKRJ24}
\end{enumerate}

\section{Results}

The following illustrates the scores, including their development over time, as well as the general properties of the submissions made by all participants during the evaluation.
The evaluation was split into two parts: first, the developers of the participating systems operated their systems themselves in an \emph{expert} session. Subsequently, additional retrieval tasks were conducted with \emph{novice} participants who had not previously seen the systems they were using.

\subsection{Scores}

\begin{figure}[!ht]
    \centering
    \includegraphics[width=0.925\linewidth]{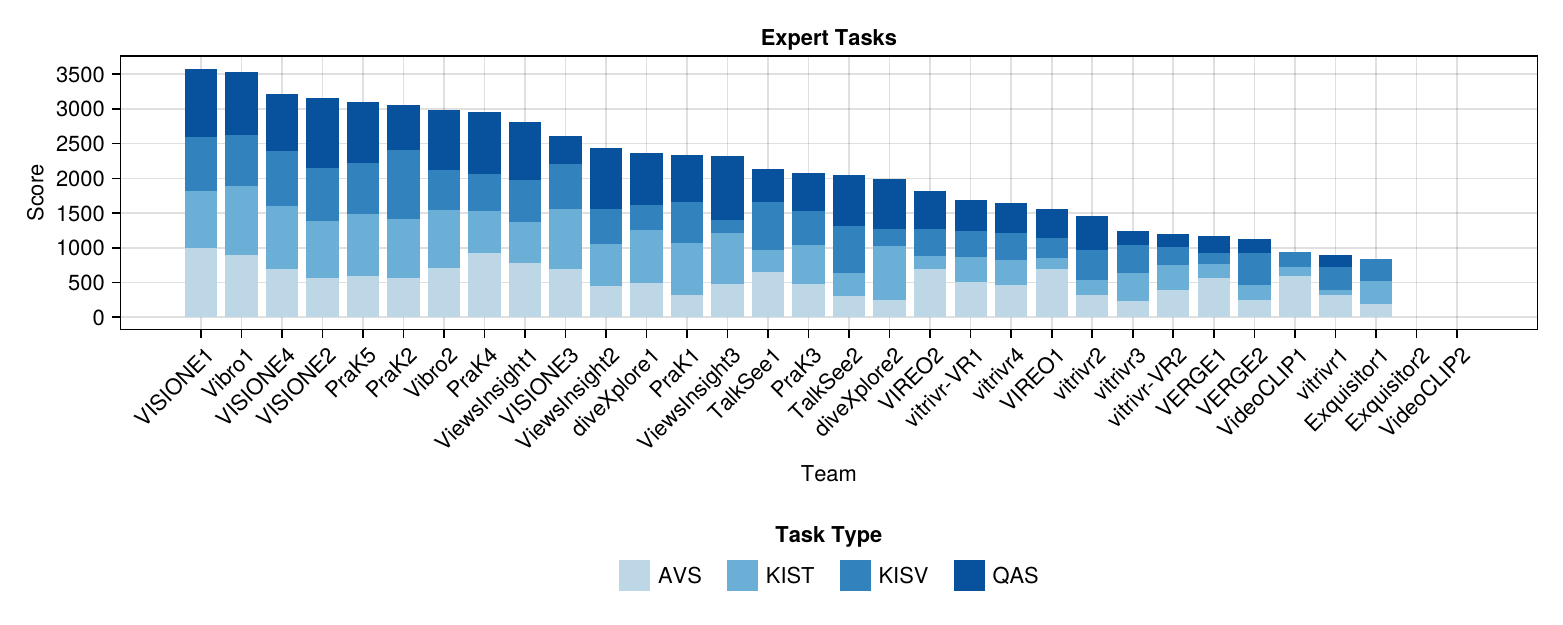}
    \caption{Scores of the expert tasks grouped by participant}
    \label{fig:scores-expert}
\end{figure}

Figure~\ref{fig:scores-expert} shows the scores of the expert tasks per participant for the four different task types: ad-hoc video search (AVS), known-item search with textual queries (KIST), known-item search with visual queries (KISV), and question answering (QAS).

\begin{figure}[!ht]
    \centering
    \includegraphics[width=0.925\linewidth]{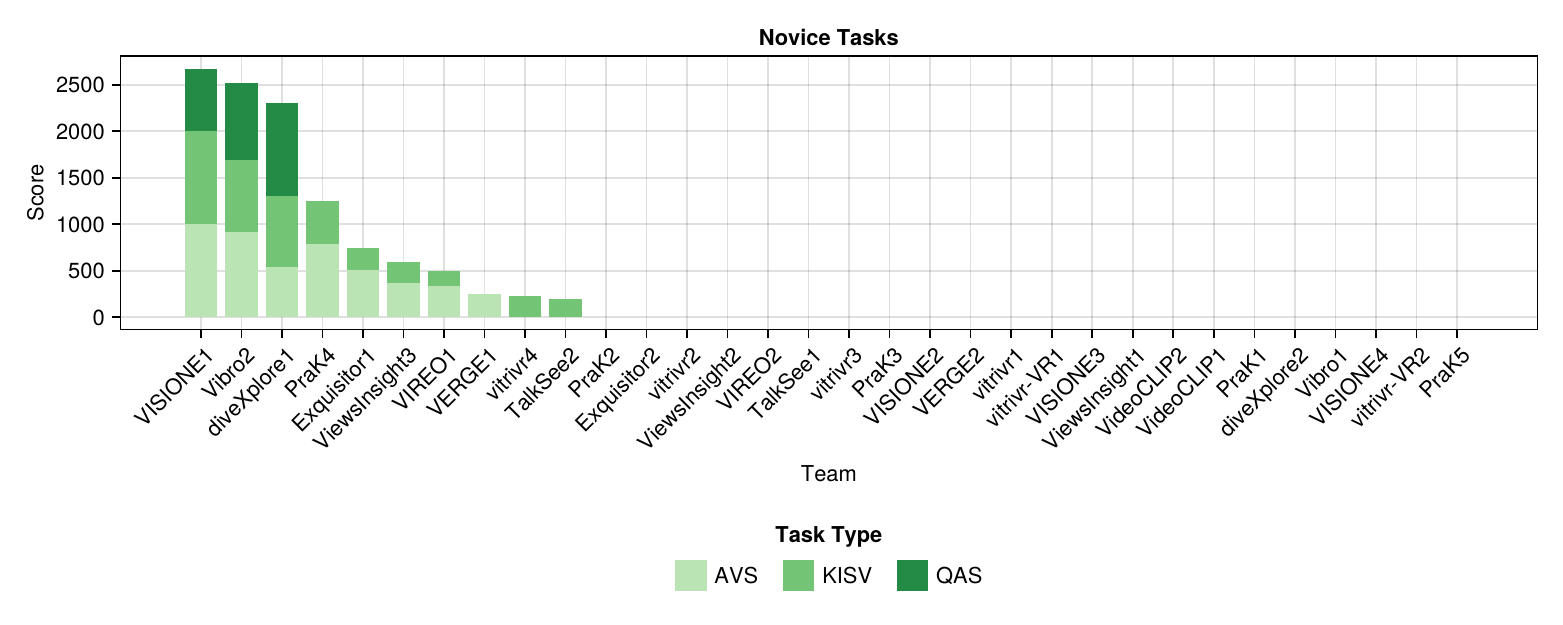}
    \caption{Scores of the novice tasks grouped by participant}
    \label{fig:scores-novice}
\end{figure}

Figure~\ref{fig:scores-novice} shows the same score distribution for the novice session. In this session, no Known-Item Search tasks with textual queries were performed.

\begin{figure}[!ht]
    \centering
    \includegraphics[width=0.925\linewidth]{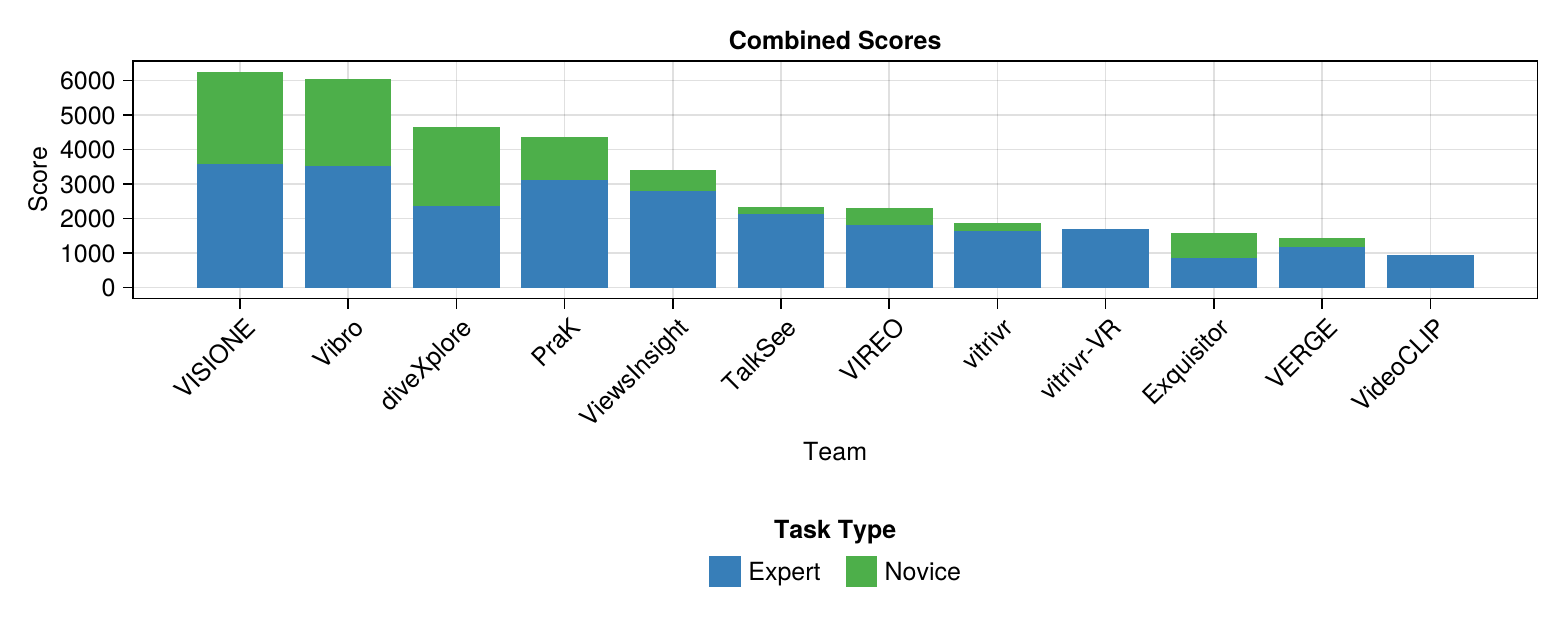}
    \caption{Combined scores, using the best-performing expert and novice per team}
    \label{fig:scores-combined}
\end{figure}

To determine the overall best-performing system/team, the highest score per system was combined from either session. The resulting final scores are shown in Figure~\ref{fig:scores-combined}.

\pagebreak

\subsection{Submissions}

\begin{figure}[!ht]
    \centering
    \includegraphics[width=\linewidth]{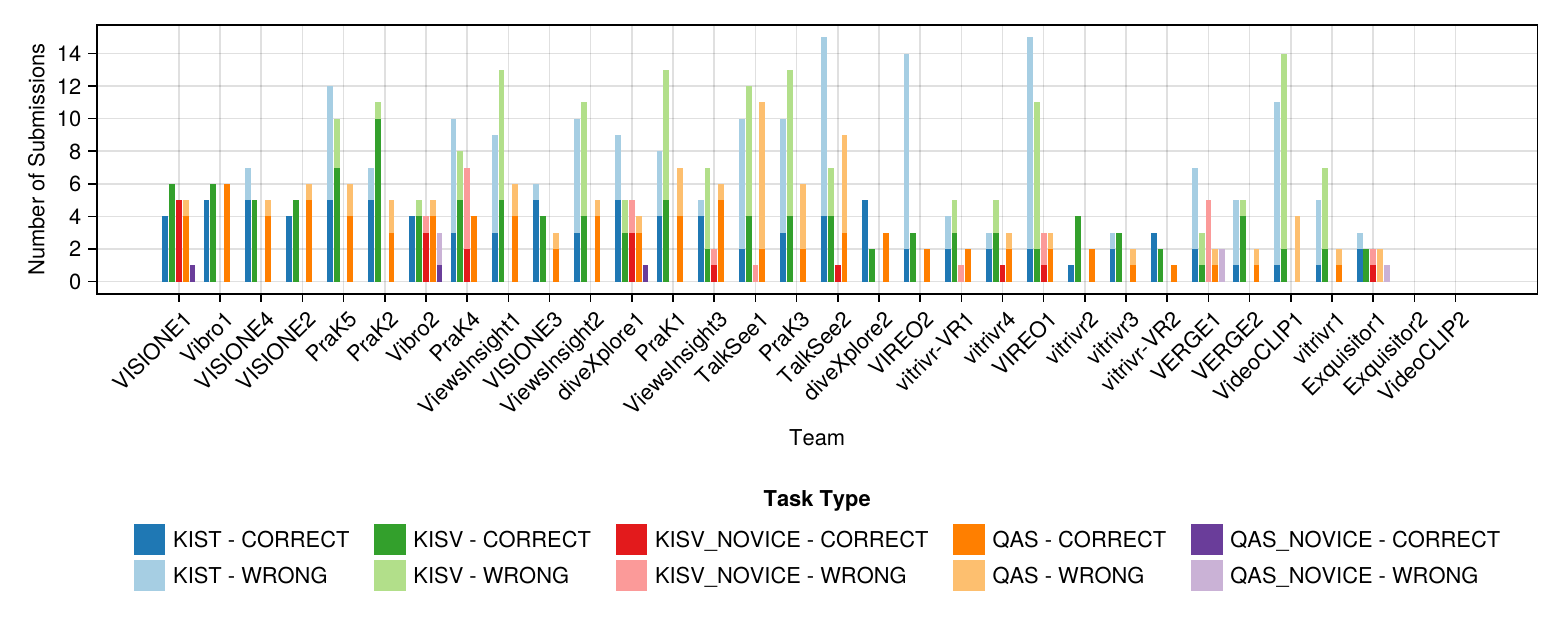}
    \caption{Number of correct and incorrect submissions per task type and participant}
    \label{fig:status}
\end{figure}

Figure~\ref{fig:status} shows the number of correct and incorrect submissions per type of task and participant. Ad-hoc search tasks are omitted due to their comparatively high number of submissions.

\begin{figure}[!ht]
    \centering
    \includegraphics[width=\linewidth]{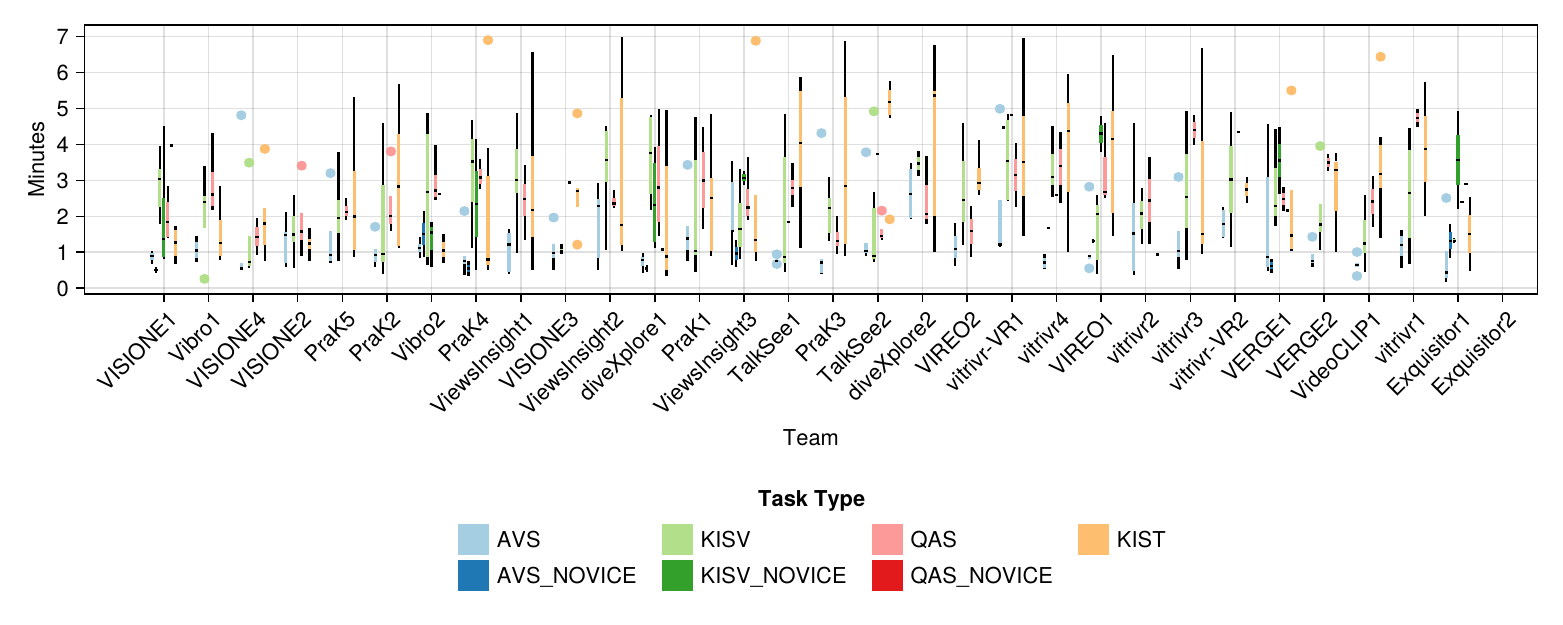}
    \caption{Distribution of the time taken (in minutes) until the first submission per task type and participant}
    \label{fig:time-to-submission}
\end{figure}

Figure~\ref{fig:time-to-submission} shows the distribution of the time taken until any participant submitted a first submission for any type of task. This serves as a rough estimate of the query processing efficiency of the different systems.

\pagebreak
\subsection{Ranking over time}

\begin{figure}[!ht]
    \centering
    \includegraphics[width=\linewidth]{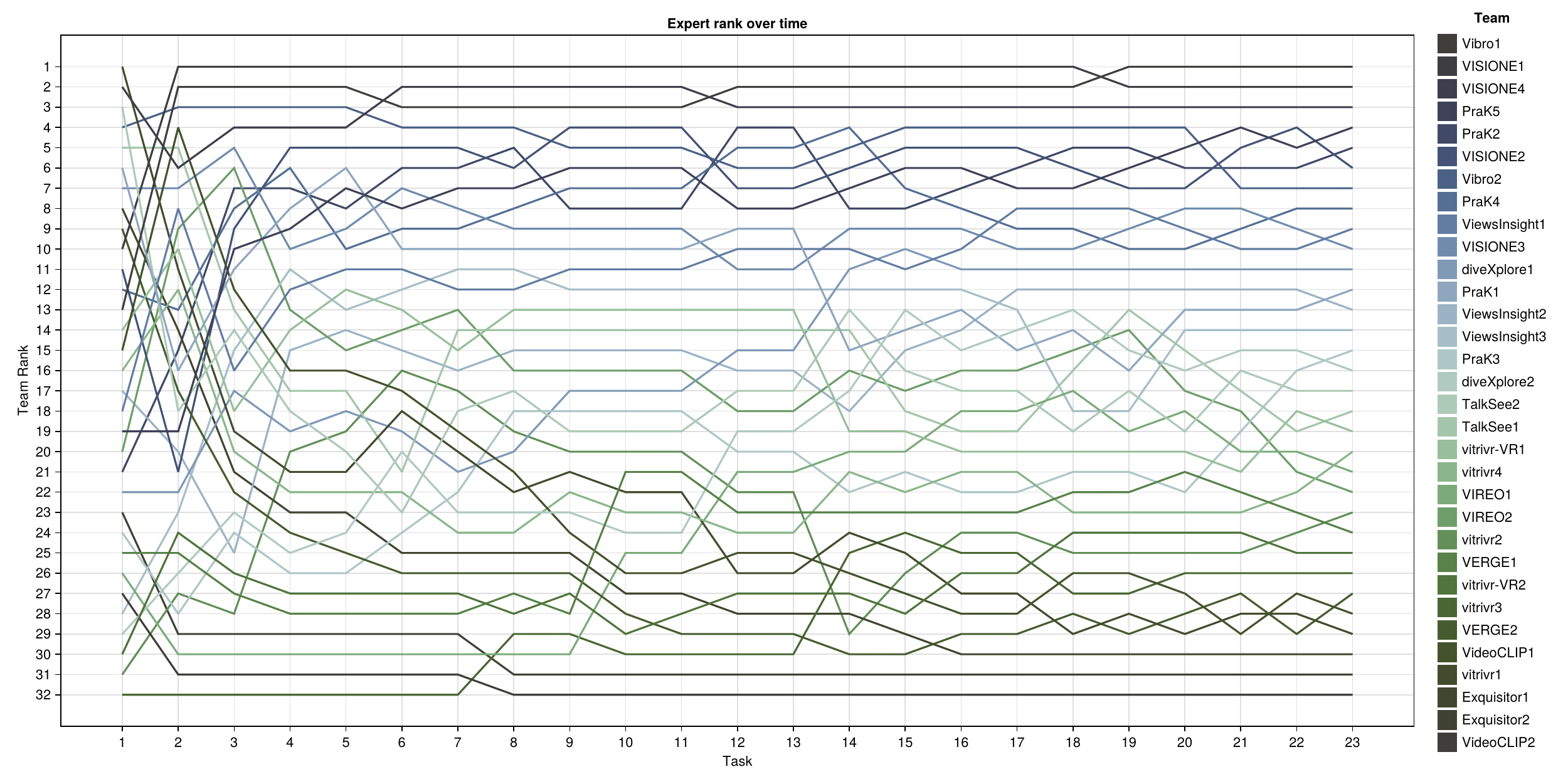}
    \caption{Raking of participants in the expert tasks over time}
    \label{fig:expert-rank-over-time}
\end{figure}

Figure~\ref{fig:expert-rank-over-time} shows the ranking of the expert participants after every task in the evaluation.

\begin{figure}[!ht]
    \centering
    \includegraphics[width=\linewidth]{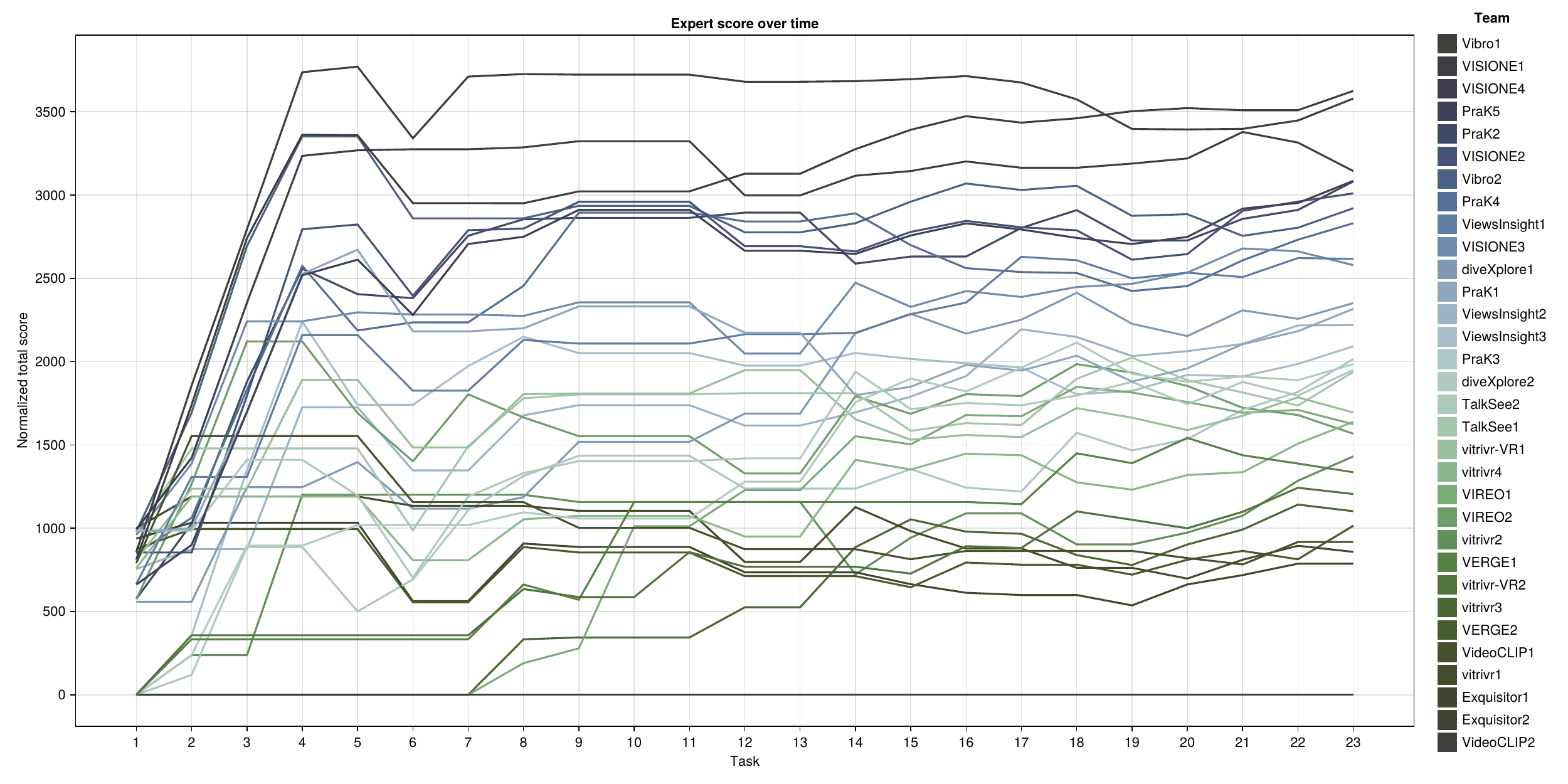}
    \caption{Score of participants in the expert tasks over time}
    \label{fig:expert-score-over-time}
\end{figure}

Figure~\ref{fig:expert-score-over-time} shows the score of the expert participants after each task in the evaluation.

\pagebreak

\begin{figure}[!ht]
    \centering
    \includegraphics[width=\linewidth]{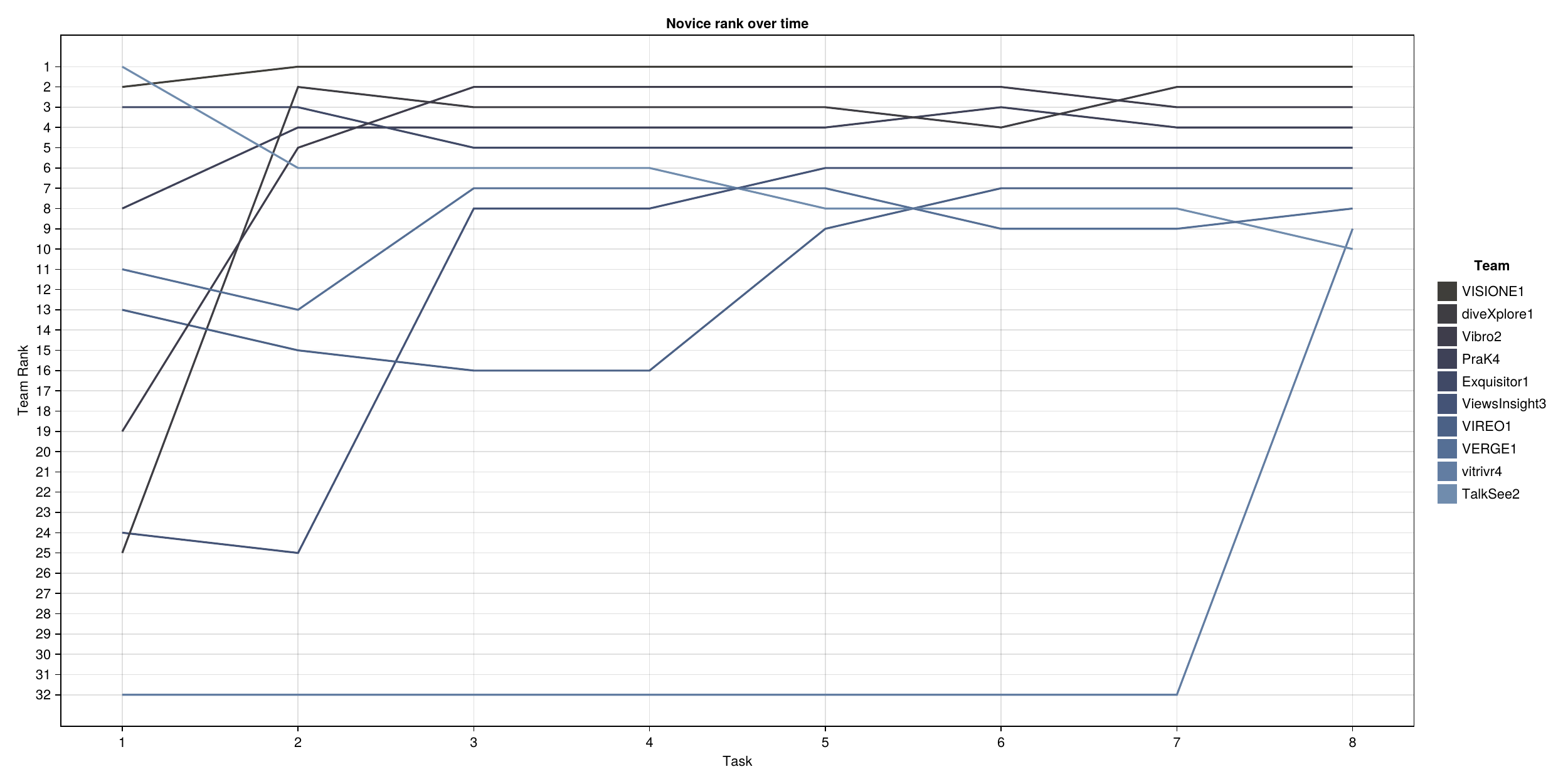}
    \caption{Raking of participants in the novice tasks over time}
    \label{fig:novice-rank-over-time}
\end{figure}

Figure~\ref{fig:novice-rank-over-time} shows the ranking of novice participants after each task in the evaluation.

\begin{figure}[!ht]
    \centering
    \includegraphics[width=\linewidth]{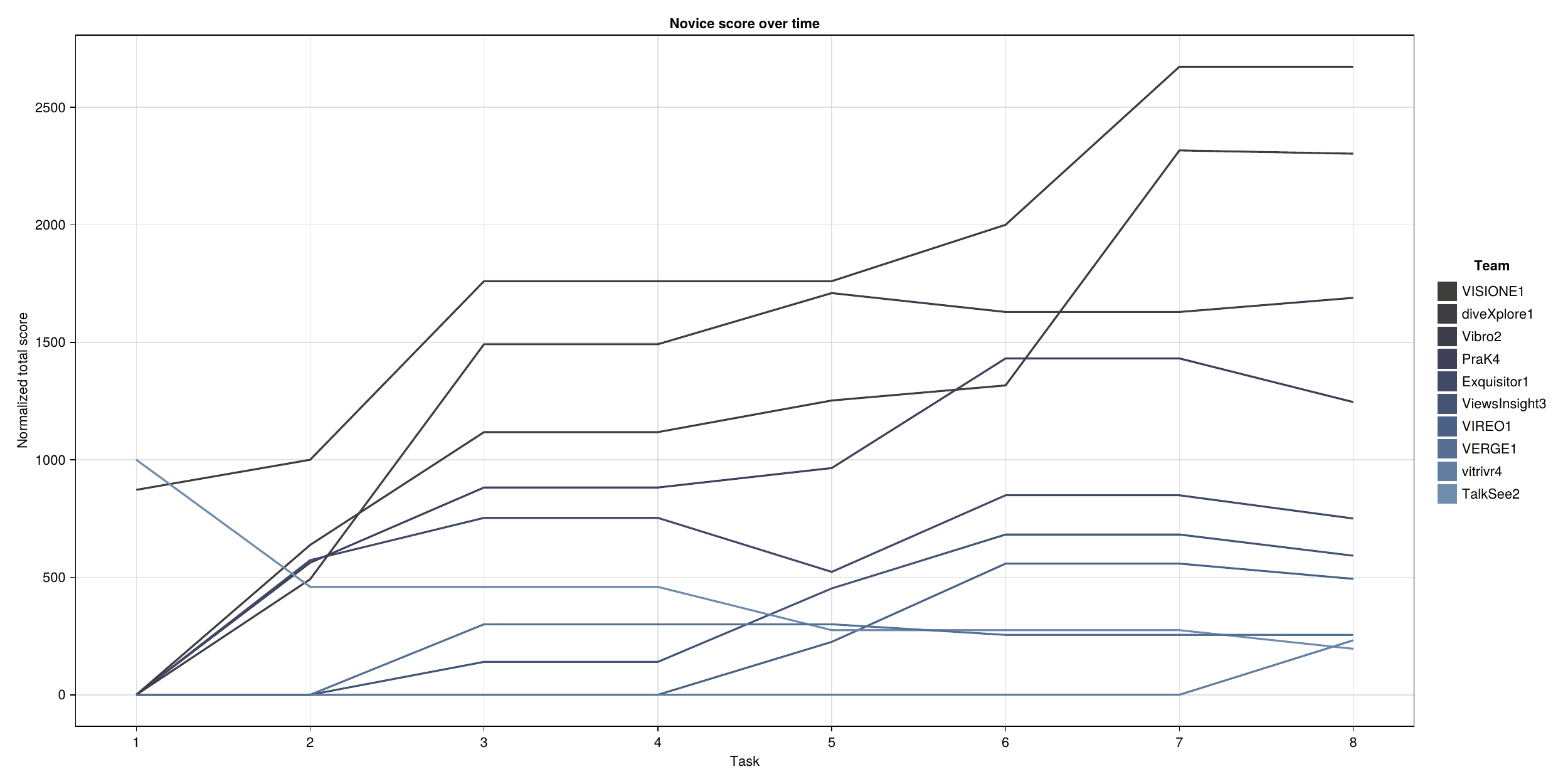}
    \caption{Score of participants in the novice tasks over time}
    \label{fig:novice-score-over-time}
\end{figure}

Figure~\ref{fig:novice-score-over-time} shows the score of novice participants after each task in the evaluation.

\pagebreak
\bibliography{bibliography}

\end{document}